\documentclass[12pt]{iopart}
\usepackage[dvips]{graphicx}
\usepackage{iopams}
\expandafter\let\csname equation*\endcsname\relax
\expandafter\let\csname endequation*\endcsname\relax
\usepackage{amsmath}
\usepackage{bm}
\usepackage{dcolumn}
\usepackage[
	bookmarks=true,
	colorlinks=true,
	urlcolor=blue,
	linkcolor=blue,
	citecolor=blue,
	pdfpagemode=UseNone,
	pdfstartview=FitH,
	pdfencoding=auto
	]{hyperref}
\usepackage[all]{hypcap}
\begin{document}

\title{Co atoms on Bi$_{2}$Se$_{3}$ revealing a coverage dependent spin reorientation transition}

\author{T~Eelbo$^1$, M~Sikora$^2$, G~Bihlmayer$^3$, M~Dobrza\'nski$^2$, A~Koz\l owski$^2$, I~Miotkowski$^4$ and R~Wiesendanger$^1$}

\address{$^1$ Institute of Applied Physics, University of Hamburg, Jungiusstra{\ss}e 11, 20355 Hamburg, Germany}
\address{$^2$ Faculty of Physics and Applied Computer Science, AGH University of Science and Technology, 30 Mickiewicza Av., 30-059 Krakow, Poland}
\address{$^3$ Peter Gr\"{u}nberg Institut and Institute for Advanced Simulation, Forschungszentrum J\"{u}lich and JARA, 52428 J\"{u}lich, Germany}
\address{$^4$ Department of Physics, Purdue University, 525 Northwestern Avenue, West Lafayette, IN, USA}

\ead{teelbo@physnet.uni-hamburg.de}

\begin{abstract}

We investigate Co nanostructures on Bi$_{2}$Se$_{3}$ by means of scanning tunneling microscopy and spectroscopy [STM/STS], X-ray absorption spectroscopy [XAS], X-ray magnetic dichroism [XMCD] and calculations using the density functional theory [DFT]. In the single adatom regime we find two different adsorption sites by STM. Our calculations reveal these to be the fcc and hcp hollow sites of the substrate. STS shows a pronounced peak for only one species of the Co adatoms indicating different electronic properties of both types. These are explained on the basis of our DFT calculations by different hybridizations with the substrate. Using XMCD we find a coverage dependent spin reorientation transition from easy-plane toward out-of-plane. We suggest clustering to be the predominant cause for this observation.

\end{abstract}


\maketitle

With decreasing dimensionality and size of nanostructures, there is an increasing importance of the electronic interactions with the supporting substrate. This may lead to new electronic and magnetic properties, e.g. on substrates with a large spin-orbit interaction [SOI] giant magnetic anisotropies of isolated Co atoms were discovered~\cite{Gambardella2003} while on alkali metals these atoms behave as being quasi-free with vanishing magnetic anisotropy energies~\cite{Gambardella2002}. The interaction of nanostructures with the supporting substrate might also induce spin reorientation transitions [SRTs]; for example, a monolayer of Fe on W(110) reveals an in-plane anisotropy while the anisotropy changes to out-of-plane if a second layer is grown on top~\cite{Pietzsch2000}. The adsorption of individual magnetic adatoms and nanostructures on exotic surfaces, like three dimensional topological insulators is currently of high interest both from a fundamental point-of-view as well as in view of potential applications. Topological insulators [TI] are a new class of materials characterized by a strong SOI which leads to gapless surface states with an odd number of Fermi-level crossings within the bulk band gap~\cite{Fu2007}. In the simplest case, a single Dirac cone consists of two spin-polarized linear dispersing branches and the crossing point, i.e. the Dirac point [DP], is protected by time reversal symmetry~\cite{Kane2005}. The time reversal symmetry is broken if magnetic impurities are introduced and the Dirac point gets massive if species with a net out-of-plane magnetic moment are added~\cite{Hor2010,Chen2010,Wray2011,Okada2011,Xu2012,Chang2013}. For this reason, we investigated the properties of Co atoms after their adsorption on the surface of Bi$_{2}$Se$_{3}$. By means of scanning tunneling microscopy and spectroscopy [STM/STS], X-ray absorption spectroscopy [XAS] and X-ray magnetic circular dichroism [XMCD] we explore the electronic and magnetic properties of the transition metal [TM] adatoms. We explain our findings based on density functional theory [DFT] calculations performed in the generalized gradient approximation~\cite{Perdew1996}.

The experiments have been carried out in two separate ultrahigh vacuum systems. Scanning tunneling microscopy and spectroscopy experiments were performed at 5K on Bi$_2$Se$_3$ single crystals \textit{in situ} cleaved at low temperatures. Using electron beam evaporators Co was directly deposited onto the cold sample at 12K to obtain well-isolated Co adatoms on the surface. To gain information about the local density of states [LDOS] electronic conductance [d$I$/d$U$] spectra were acquired by means of a lock-in technique using a modulation voltage $U_{\rm{mod}}=20$mV and a frequency $f=5$kHz. The XAS and XMCD experiments have been carried out at the ID08 beamline at the European Synchrotron Radiation Facility. While the measurement temperature was about $T\approx 8$K, the single crystals had to be cleaved at room temperature and immediately cooled down afterward. Co atoms have been deposited by using an electron beam evaporator with the substrate remaining in the measurement stage at a temperature of $T\approx 10$K. X-ray absorption spectra were obtained in the total-electron-yield mode using almost $100 \%$ polarized light. Magnetic fields of up to 5T were applied collinear to the incident beam. In addition, the sample was rotated from normal [$0^\circ$] to steep [$70^\circ$] incidence angle to obtain information about the in- and out-of-plane magnetic properties. All spectra have been normalized with respect to the incident beam intensity and the Co $L_3$ pre-edge intensity.

\begin{figure}
\begin{center}
\includegraphics[width=0.5\textwidth]{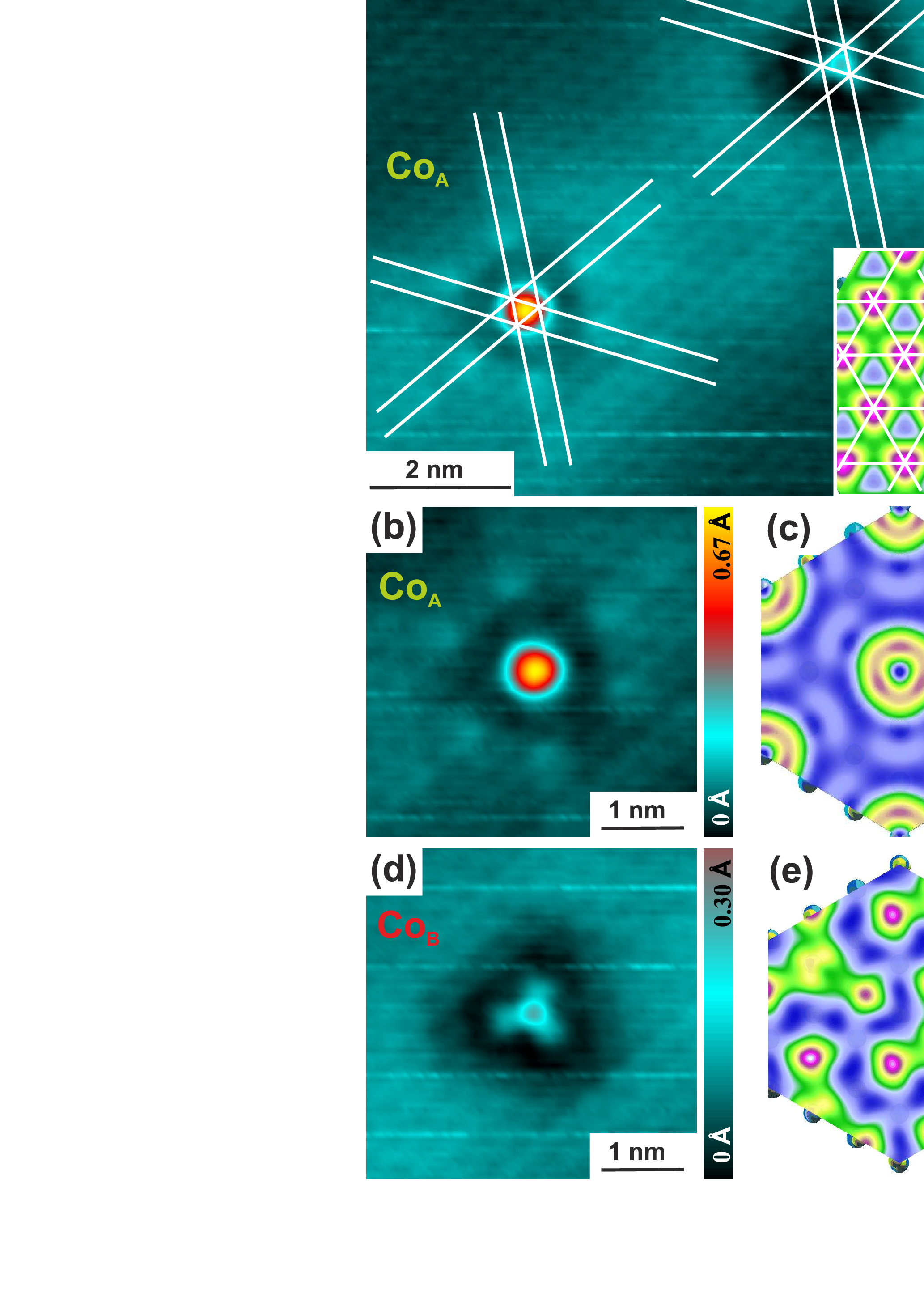}
\caption{\label{Fig1} (a) STM topography of two different isolated Co adatoms on Bi$_{2}$Se$_{3}$. The tunneling parameters are $U=0.2$V and $I=0.75$nA. The inset shows a theoretical simulation of the bare surface. The white lines are guide lines to the eyes representing Se top sites. (b) and (d) Magnified views on both types of adsorbates showing the affected surface in their vicinity as well. (c) and (e) Simulations of Co atoms adsorbed in the fcc and hcp hollow site for a bias voltage of $U=0.1$V.}
\end{center}
\end{figure}

We used STM to address the local properties of the Co adatoms. Contrary to a recent work on Co monomers on Bi$_{2}$Se$_{3}$, where the authors concluded Co atoms to adsorb on Se top sites~\cite{Ye2012}, we find two different types of Co atoms on Bi$_{2}$Se$_{3}$. Figure~\ref{Fig1}(a) proves that both species [in the following Co$_{\rm A/B}$] occupy different adsorption sites which we illustrate by the white guidelines to the eyes. Regarding their apparent heights and shapes as well as the influence on the surrounding substrate further differences appear among both species. On the one hand, for the stabilization voltage chosen, Co$_{\rm A}$ shows a larger apparent height [$\approx0.7$\AA]~than Co$_{\rm B}$ [$\approx0.3$\AA]. On the other hand, in case of Co$_{\rm A}$, the substrate shows an almost sixfold symmetric pattern in its vicinity whereas Co$_{\rm B}$ induces a threefold symmetric pattern, compare figures~\ref{Fig1}(b) and (d). Importantly, at no bias voltage, the Co adatoms appear as dark triangular depressions which generally would hint toward a substitution of Bi at its lattice site, which e.g. has been found for Fe adatoms after room temperature annealing~\cite{Schlenk2013}. For a coverage of 0.01 monolayer equivalent [MLE] the relative abundance of both species is approximately three to one with a predominance of Co$_{\rm A}$ type adatoms indicating that the adsorption site of Co$_{\rm A}$ is energetically favorable. We note that depending on the tunneling parameters [e.g. during STS] Co$_{\rm B}$ type atoms can be manipulated and afterward appear as Co$_{\rm A}$. A change from Co$_{\rm A}$ toward Co$_{\rm B}$ has never been observed.

To elucidate our STM/STS observations, DFT calculations using the full-potential linearized augmented planewave method as implemented in the {\sc Fleur}-code~\cite{fleur} have been performed. The model comprises a $\left(\sqrt{3} \times \sqrt{3}\right)\rm{R}30^{\circ}$ unit cell of four quintuple layers of Bi$_{2}$Se$_{3}$. We find the fcc hollow position to be the energetically most favorable adsorption site. Nevertheless, the hcp hollow position is unfavorable by only $\approx90$meV per atom. For a comparison with the experimental data, STM topographies of the bare surface have been simulated using the vacuum density of states up to +100mV, compare the inset of figure~\ref{Fig1}(a). At this voltage the Se atoms show up as bright triangles. Guidelines which cross at these positions reveal that Co$_{\rm A/B}$ do occupy different hollow sites. Further simulations of Co atoms adsorbed in the fcc and hcp hollow sites reveal different appearances of both species, shown in figures~\ref{Fig1}(c) and (e). The good agreement between the simulated topographies and the experimental data let us conclude that Co$_{\rm A}$ is adsorbed in the fcc hollow site while Co$_{\rm B}$ is adsorbed in the hcp hollow site. This assignment is further supported by the relative abundance if the energy difference between both adsorption sites is taken into consideration. 

\begin{figure}[b]
\begin{center}
\includegraphics[width=0.55\textwidth]{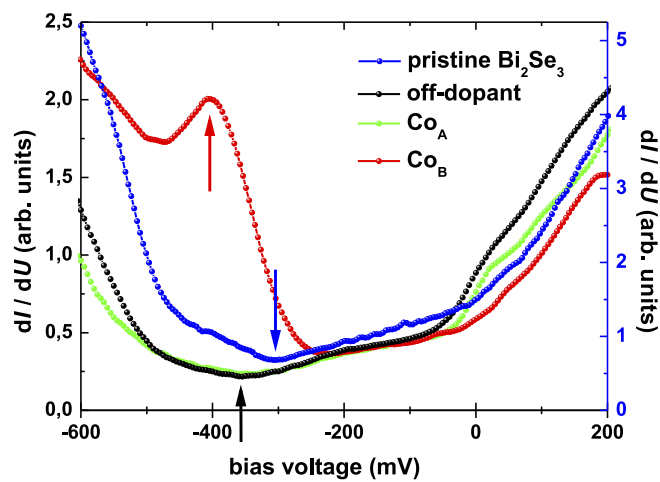}
\caption{\label{Fig2} STS of pristine Bi$_{2}$Se$_{3}$ [right y-axis] and 0.01 MLE Co/Bi$_{2}$Se$_{3}$ [left y-axis]. The off-dopant spectrum has been acquired after Co deposition far away from any Co adatom. The blue and black arrows indicate the energetic position of the DP before and after Co deposition and, hence, reveal a shift of $\approx-50$mV. Tunneling parameters are $U=0.2$V and $I=0.1$nA.}
\end{center}
\end{figure}

\begin{figure}
\begin{center}
\includegraphics[width=0.8\textwidth]{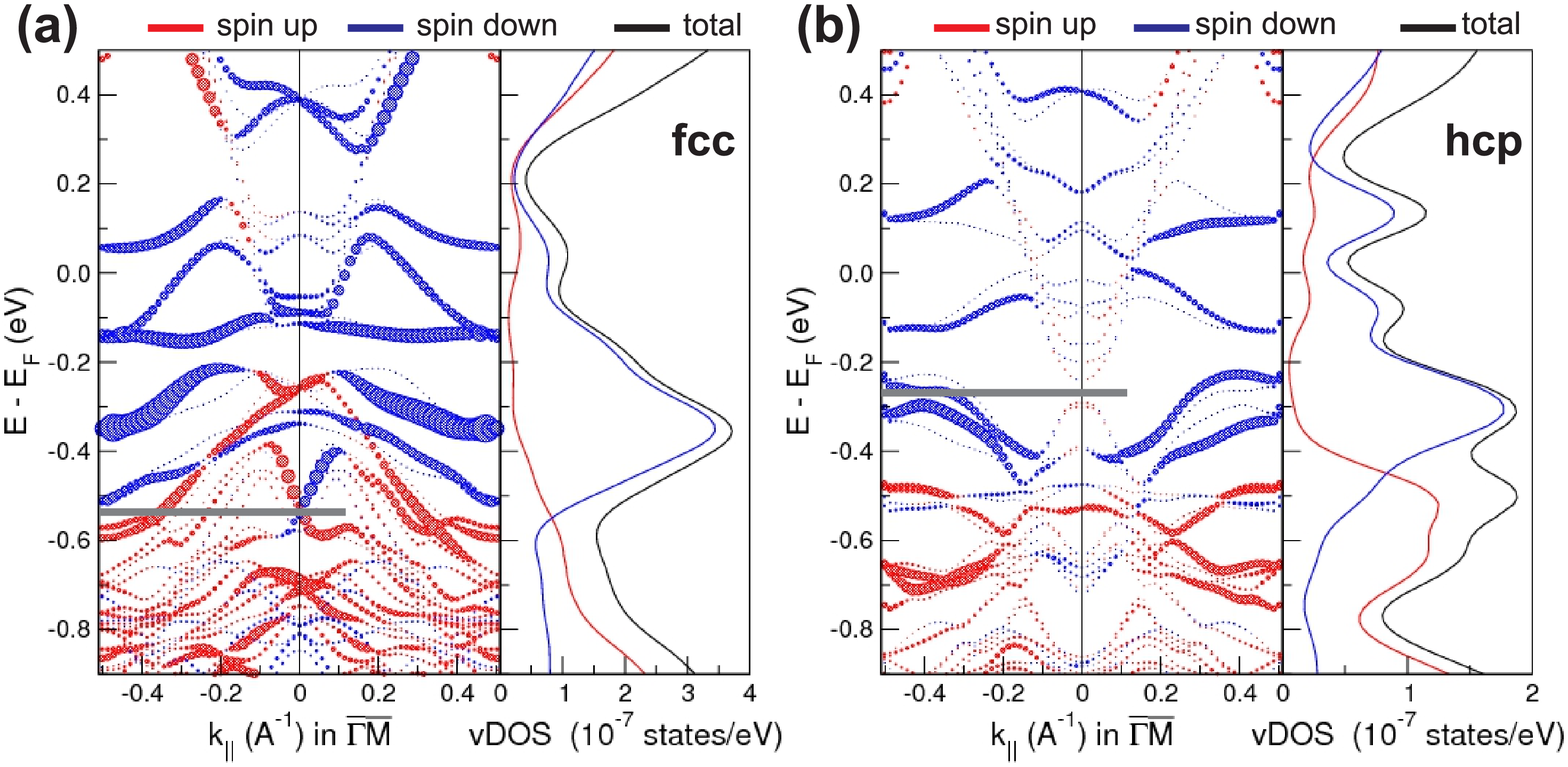}\\
\caption{\label{Fig3} Fat band analysis and vacuum DOS of Co adsorbed in the fcc hollow site (a) and in the hcp hollow site (b). The red and blue lines indicate spin up and spin down states with respect to the easy axes while the black lines depict the total DOS. The size of the circles gives the spin-polarization of the states in a region above the surface. The gray lines depict the computed Dirac points at -0.55eV and -0.25eV for fcc and hcp occupation, respectively.}
\end{center}
\end{figure}

In addition, the simulations indicate a relaxation of the Co atoms into the surface of Bi$_{2}$Se$_{3}$ in both cases by $\approx0.2$\AA. Opposite to theory, the experimentally resolved apparent heights differ significantly in case of fcc and hcp occupation. Thus, we relate the difference of the observed apparent heights to different hybridizations with the surrounding atoms resulting in different electronic properties for both species [adsorption sites]. Therefore, the adatoms have been investigated by means of STS. Figure~\ref{Fig2} shows STS spectra of pristine Bi$_{2}$Se$_{3}$ and Co/Bi$_{2}$Se$_{3}$. While for the pristine crystal the onset of the bulk valence [conduction] band is detected at $\approx-450$mV [$\approx0$mV], we find a global minimum [blue arrow] which we assign to the Dirac point [DP] at $-300$mV. The shift with respect to the Fermi level indicates that the crystal is naturally electron doped, being in agreement with previous observations~\cite{Hor2009,Bianchi2010} and attributed to the existence of Se$_{\text{Bi}}$ antisite defects~\cite{Urazhdin2002}. Upon Co deposition of 0.01 MLE the off-dopant spectrum exhibits a global minimum [black arrow] at $\approx-350$mV, which hence depicts an additional shift of $\approx-50$mV of the DP with respect to the Fermi level. We conclude that Co further n-dopes the substrate and acts as a donor. In contrast to recent predictions~\cite{Liu2009,Schmidt2011}, no indication of a global surface band gap has been found after the deposition of Co adatoms. Regarding the STS spectra of the adatoms, no distinct resonances have been found in case of Co$_{\rm A}$, whereas Co$_{\rm B}$ type adatoms reveal a pronounced peak at $-400$mV. According to our theoretical calculations, the different electronic properties and the appearance of a resonance only in the hcp case are caused by different hybridizations of the Co $3d$ electrons with Bi$_{2}$Se$_{3}$, as is illustrated in figure~\ref{Fig3}. The fat band analysis of Co adatoms in the fcc [figure~\ref{Fig3}(a)] and hcp hollow site [figure~\ref{Fig3}(b)] reveal the Dirac points to be located at -0.55eV and -0.25eV, respectively. Spin up and spin down weights are defined with respect to the easy axes of both configurations, which are in-plane for fcc and out-of-plane for the hcp adsorption. While in case of hcp occupation the band structure shows two additional Co bands at $\approx-0.1$eV below the DP, these bands are shifted toward $\approx0.3$eV above the DP in case of fcc. We also notice the different dispersion of these bands in the fcc and hcp geometry: while in the latter case the hybridization with the topmost valence band (mainly Bi $p_z$ states~\cite{Zhang2009}) leads to an increase of the binding energy at the center of the Brillouin-zone, in the fcc-case the dispersion of the band is inverted. As a result, the Co-induced peak in the vacuum DOS is energetically higher in the hcp than in the fcc case. Although the calculations were performed for a coverage of 0.33 MLE and, hence, the absolute positions of the peaks cannot be directly compared to the STS data, the differences between the fcc and hcp occupation are significant and can be related to the different STS observations, in particular the pronounced peak observed for Co$_{\rm B}$.
 
\begin{figure}[b]
\begin{center}
\includegraphics[width=0.7\textwidth]{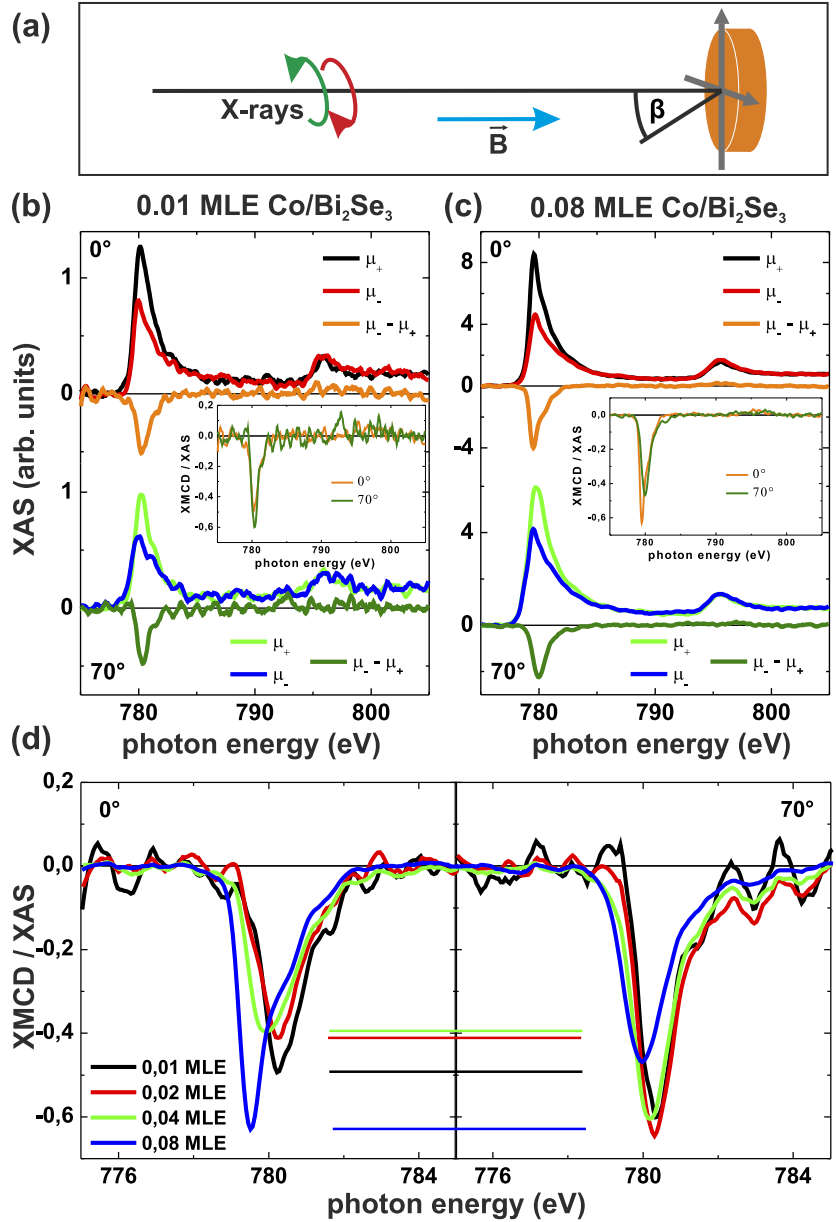}\\
\caption{\label{Fig4} (a) Sketch of the experimental setup. The magnetic field can be applied parallel to the incident beam direction whereas the sample can be inclined with respect to this direction. (b) XAS and XMCD spectra for 0.01 MLE of Co/Bi$_{2}$Se$_{3}$ for normal (upper panel) and grazing (lower panel) incidence angle. The inset shows the XMCD signal strength normalized by the XAS $L_3$ peak height. (c) Spectra for an increased coverage of 0.08 MLE. (d) Normalized XMCD signals for a series of coverages ranging between (b) and (c) for normal (left panel) and grazing (right panel) incidence angle. The colored lines are guidelines to the eyes indicating differences between both angles.}
\end{center}
\end{figure}

The electronic and magnetic properties have been further tested by XAS and XMCD measurements, summarized in figure~\ref{Fig4}. A sketch of the experimental setup is depicted in figure~\ref{Fig4}(a). Different coverages ranging from 0.01 MLE to 0.08 MLE have been investigated. Independent of the coverage, the shapes of the XAS spectra show no distinct multipeak structures besides a slight shoulder on the high-energy side of the Co $L_3$ edge at approximately $780.2$eV, compare figures~\ref{Fig4}(b) and (c). The XAS line shape suggests the Co atoms to be in the electronic configuration of $3d^7$~\cite{Laan1992}. This result is in agreement to recent works on Fe/Bi$_{2}$Se$_{3}$~\cite{Honolka2012} as well as Fe and Co/Bi$_{2}$Te$_{3}$~\cite{Shelford2012} where the TM adatoms were found to be in their pristine configuration as well. Furthermore, the XAS spectra can be used to estimate the branching ratio [$BR$] which serves as an indicator of the spin character of the ground state of the adatoms~\cite{branching,Thole1988}. Within the coverage range investigated, we find a constant value of $BR=0.84 \pm 0.01$ which suggests a high-spin ground state of the Co adsorbates. The inset in figure~\ref{Fig4}(b) shows the angular dependence of the XMCD signal normalized with respect to the $L_3$ XAS intensity, which can be used as an indicator for the easy axis of the Co adatoms. For the low coverage regime we find the signal strength for normal incidence angle to be enhanced by about 20\% compared to the signal at grazing incidence angle. This suggests the easy axis to reside in the surface plane similar to Fe/Bi$_{2}$Se$_{3}$~\cite{Honolka2012} and contrary to predictions of an out-of-plane anisotropy for Co/Bi$_{2}$Se$_{3}$ in~\cite{Ye2012, Schmidt2011}. The experimentally indicated easy-plane easy axis is particularly in line with our calculations, which predict an easy-plane magnetocrystalline anisotropy energy [MAE] in case of fcc hollow site occupation [Co$_{\rm A}$; $K_{fcc}=-6$meV] and an out-of-plane MAE in case of the hcp hollow site [Co$_{\rm B}$; $K_{hcp}=+3$meV], where $K$ denotes the MAE per adatom. Taking the relative abundance into account [low coverage: $n_{\rm{Co}_{\rm A}} / n_{\rm{Co}_{\rm B}} = 3/1$] an easy-plane anisotropy is theoretically expected. The agreement further supports the assignment of the adsorption sites to the different types of Co adatoms. We note, that a site-dependent anisotropy has been reported before on metal substrates~\cite{Blonski2010, Khajetoorians2013} and as usual depends on the layer thickness, which -- for our calculations -- has been 0.33 MLE. Moreover, we investigated the anisotropy using the normalized $L_3$ XMCD intensity as a function of the coverage, see figure~\ref{Fig4}(d). Based on the accuracy level given for the estimation of the coverage, the relative $L_3$ XMCD/XAS intensities acquired at both angles suggest an in-plane easy axis at low coverages [$0.01$ MLE -- $0.04$ MLE] and an out-of-plane easy axis for 0.08 MLE.

Unfortunately, the magnetic moments of the Co adatoms have not been saturated at the maximum magnetic field available [5T] and, thus, deducing the orbital and effective spin moments would be unreliable. However, the ratio [$R$] of orbital to effective spin moment~\cite{ratio} is independent of the saturation and can be used for conclusions about the investigated sample. In the regime of single atoms, we find $R=0.33 \pm 0.02$, which is in good agreement with~\cite{Ye2012} and significantly larger than the Co bulk value~\cite{Chen1995}. The increase of the Co coverage toward 0.08 MLE goes hand in hand with a gain of the ratio until $R=0.49 \pm 0.03$ which indicates relevant changes in the orbital and effective spin moments. Such a trend was observed before~\cite{Ye2012} although the orbital moment is expected to diminish upon increasing the mean cluster size~\cite{Gambardella2003}. This contradiction can be understood taking the SRT into consideration. In case of the low coverage [0.01 MLE] we find an easy-plane anisotropy, i.e. $m_L^{\text{x}}>m_L^{\text{z}}$ for z denoting the direction parallel to the surface normal. Hence, the ratio is given by $R^{0.01}=m_L^{\text{x}}/m_S^{\text{x}}=0.33 \pm 0.02$. For the high coverage regime [0.08 MLE] we deduce an out-of-plane easy axis which, according to Bruno~\cite{Bruno1989}, means that $m_L^{\text{x}}<m_L^{\text{z}}$ in this case. Furthermore, in good agreement with our theoretical model, we can assume the magnetic spin moment to be independent of the orientation, i.e. $m_S^{\text{x}}\approx m_S^{\text{z}}$ [$1.17 \mu_{\rm B}/\text{atom}\approx 1.15 \mu_{\rm B}/\text{atom}$]. If the ratio of in-plane orbital moment and in-plane spin moment additionally remains constant while increasing the coverage, then an increase of the ratio given by $R^{0.08}=m_L^{\text{z}}/m_S^{\text{z}}=0.49 \pm 0.03$ is possible and essentially driven by the SRT. We note, that these conclusions are based on the assumption of a vanishing spin dipole moment $m_D$ and that the magnetic spin moments have been calculated for a coverage of 0.33 MLE.

In order to understand the spin reorientation transition in more detail, we performed a series of STM/STS experiments at elevated coverages [up to 0.1 MLE] and determined the ratio of adatoms showing a resonance at -400mV, i.e. Co$_{\rm B}$, to those not exhibiting this resonance. Although we observe slight changes of this ratio, no drastic modifications were found. Therefore, we experimentally rule out the possibility that a change of the relative population of fcc and hcp hollow sites upon increasing the total Co coverage might cause the observed SRT. Instead, we assign the transition to the decrease of the mean distance between the adatoms upon increasing the Co coverage. From this an emerging interaction as well as the growth of clusters seems plausible, which consequently causes serious modifications of the electronic properties as well as the anisotropy of the magnetic moments. This conclusion is supported by the XAS line shape showing a decrease of the high-energy shoulder at the Co $L_3$ edge upon increasing the coverage, compare figures~\ref{Fig4}(b) and (c). Furthermore, the series of coverage dependent XMCD spectra [figure~\ref{Fig4}(d)] shows that the peak position monotonically shifts downward in energy while increasing the coverage. The exact shape of the XAS spectra is determined by the average chemical state and the average crystal field of the ensemble probed by the X-ray beam, since the symmetry and the splitting of the crystal field as well as the chemical state might be slightly different for hcp and fcc occupation. Hence, we relate the variations of the XAS line shape and the shift of the $L_3$ peak position to changes of these quantities. The upcoming growth of clusters while the coverage is raised certainly influences both and, therefore, is a likely explanation for the mutation of the XAS/XMCD spectra. Note, that in the high coverage regime, similar to the low coverage regime, no gap opening has been detected for spectra acquired as far away from any impurity as possible, although the out-of-plane easy axis is evidenced by XMCD. A possible reason might be given in view of the clusters being the predominant cause for the variation of the anisotropy, since spectra in their vicinity have not been measurable because the tunneling conditions have not been sufficiently stable at these locations. In contrast, by the global technique of XMCD the clusters' influence can be easily detected. In general, our conclusions are supported by a recent study on bulk-doped Mn-Bi$_2$Se$_3$ where excess Mn clusters have been observed on the crystal's surface and suggested to significantly influence the surface magnetization of the sample~\cite{Zhang2012}.

In conclusion, by means of scanning tunneling microscopy we find Co adatoms to occupy fcc and hcp hollow sites on Bi$_{2}$Se$_{3}$ after low temperature deposition. The irreversible switching from Co$_{\rm B}$ into Co$_{\rm A}$ as well as the employed DFT calculations let us assign the Co$_{\rm A}$ type adatoms to be adsorbed in fcc hollow sites. The DFT calculations further indicate different hybridization effects to be the cause for different electronic properties of both species found within the STS spectra. Using XMCD, we determine an easy axis of the magnetization in the surface plane for the low coverage regime, whereas it changes to out-of-plane upon increasing the Co coverage. Cluster formation is suspected to be responsible for the observed spin reorientation transition. 

We thank M.~Gyamfi for fruitful discussions. Financial support from the ERC Advanced Grant FURORE and the DFG-Sonderforschungsbereich 668 is gratefully acknowledged. We gratefully acknowledge computing time from the Jülich supercomputing centre (JSC).


\section*{References}

\begin{thebibliography}{23}

\bibitem{Gambardella2003} Gambardella P \etal 2003 {\it Science} {\bf 300} 1130

\bibitem{Gambardella2002} Gambardella P, Dhesi S S, Gardonio S, Grazioli C, Ohresser P and Carbone C 2002 {\it Phys. Rev. Lett.} {\bf 88} 047202

\bibitem{Pietzsch2000} Pietzsch O, Kubetzka A, Bode M and Wiesendanger R 2000 {\it Phys. Rev. Lett.} {\bf 84} 5212

\bibitem{Fu2007} Fu L, Kane C L and Mele E J 2007 {\it Phys. Rev. Lett.} {\bf 98} 106803

\bibitem{Kane2005} Kane C L and Mele E J 2005 {\it Phys. Rev. Lett.} {\bf 95} 226801

\bibitem{Hor2010} Hor Y S \etal 2010 {\it Phys. Rev. B} {\bf 81} 195203

\bibitem{Chen2010} Chen Y L \etal 2010 {\it Science} {\bf 329} 659

\bibitem{Wray2011} Wray L A, Xu S Y, Xia Y, Hsieh D, Fedorov A V, Hor Y S, Cava R J, Bansil A, Lin H and Hasan M Z 2011 {\it Nat. Phys.} {\bf 7} 32

\bibitem{Okada2011} Okada Y \etal 2011 {\it Phys. Rev. Lett.} {\bf 106} 206805

\bibitem{Xu2012} Xu S-Y \etal 2012 {\it Nat. Phys.} {\bf 8} 616

\bibitem{Chang2013} Chang C-Z \etal 2013 {\it Science} {\bf 340} 167

\bibitem{Perdew1996} Perdew J P, Burke K and Ernzerhof M 1996 {\it Phys. Rev. Lett.} {\bf 77} 3865

\bibitem{Ye2012} Ye M \etal 2012 {\it Phys. Rev. B} {\bf 85} 205317 

\bibitem{Schlenk2013} Schlenk T \etal 2013 {\it Phys. Rev. Lett.} {\bf 110} 126804

\bibitem{fleur} For a detailed description, see http://www.flapw.de

\bibitem{Hor2009} Hor Y S, Richardella A, Roushan P, Xia Y, Checkelsky J G, Yazdani A, Hasan M Z, Ong N P and Cava R J 2009 {\it Phys. Rev. B} {\bf 79} 195208 

\bibitem{Bianchi2010} Bianchi M, Guan D, Bao S, Mi J, {Brummerstedt Iversen} B, King P D C and Hofmann P 2010 {\it Nat. Commun.} {\bf 1} 128

\bibitem{Urazhdin2002} Urazhdin S, Bilc D, Tessmer S H, Mahanti S D, Kyratsi T and Kanatzidis M G 2002 {\it Phys. Rev. B} {\bf 66} 161306(R)

\bibitem{Liu2009} Liu Q, Liu C X, Xu C, Qi X L, and Zhang S C 2009 {\it Phys. Rev. Lett.} {\bf 102} 156603

\bibitem{Schmidt2011} Schmidt T M, Miwa R H and Fazzio A 2011 {\it Phys. Rev. B} {\bf 84} 245418

\bibitem{Zhang2009} Zhang H, Liu C X, Qi X L, Dai X, Fang Z and Zhang S C 2009 {\it Nat. Phys.} {\bf 5} 438

\bibitem{Laan1992} {van der Laan} G and Kirkman I W 1992 {\it J. Phys.: Condens. Matter} {\bf 4} 4189

\bibitem{Honolka2012} Honolka J \etal 2012 {\it Phys. Rev. Lett.} {\bf 108} 256811

\bibitem{Shelford2012} Shelford L R, Hesjedal T, {Collins-McIntyre} L, Dhesi S S, Maccherozzi F and {van der Laan} G 2012 {\it Phys. Rev. B} {\bf 86} 081304(R)

\bibitem{branching} $BR$: Ratio of the integrated XAS of the $L_3$ edge divided by the sum of the integrated XAS of the $L_3$ and $L_2$ edges.

\bibitem{Thole1988} Thole B T and van~der~Laan G 1988 {\it Phys. Rev. B} {\bf 38} 3158

\bibitem{Blonski2010} B{\l}o\'{n}ski P, Lehnert A, Dennler S, Rusponi S, Etzkorn M, Moulas G, Bencok P, Gambardella P, Brune H and Hafner J 2010 {\it Phys. Rev. B} {\bf 81} 104426

\bibitem{Khajetoorians2013} Khajetoorians A A, Schlenk T, Schweflinghaus B, dos Santos Dias M, Steinbrecher M, Bouhassoune M, Lounis S, Wiebe J and Wiesendanger R 2013 {\it Phys. Rev. Lett.} {\bf 111} 157204

\bibitem{ratio} $R$: Ratio of orbital magnetic moment ($m_{L}$) to the sum of spin moment ($m_{S}$) and spin dipole moment ($m_{D}$) according to $R$~=~$m_{L}$~/~($m_{S}$~+~7~$m_{D}$), where the moments are given in $\mu_{\rm B}$/atom and are projected along the incident beam direction.

\bibitem{Chen1995} Chen C T, Idzerda Y U, Lin H J, Smith N V, Meigs G, Chaban E, Ho G H, Pellegrin E and Sette F 1995 {\it Phys. Rev. Lett.} {\bf 75} 152

\bibitem{Bruno1989} Bruno P 1989 {\it Phys. Rev. B} {\bf 39} 865

\bibitem{Zhang2012} Zhang D \etal 2012 {\it Phys. Rev. B} {\bf 86} 205127

\end{thebibliography}
\end{document}